\documentclass[a4paper,fleqn,usenatbib,useAMS]{mnras}
\usepackage{graphicx}	
\usepackage{amsmath}	
\usepackage{amssymb}	
\usepackage{multicol}        
\usepackage{bm}		
\usepackage{pdflscape}	

\usepackage[T1]{fontenc}
\usepackage{ae,aecompl}

\pdfoutput=1

\def\beq{\begin{equation}}
\def\eeq{\end{equation}}
\def\bey{\begin{eqnarray}}
\def\eey{\end{eqnarray}}

\title[Energetic constraints on electromagnetic signals from double black hole mergers]{Energetic constraints on electromagnetic signals from double black hole mergers}

\author[Dai et al.]{
Lixin Dai,$^{1,3}$\thanks{E-mail: cosimo@umd.edu (LD)}
Jonathan C. McKinney,$^{1,3}$
and M. Coleman Miller$^{2,3}$
\\
$^{1}$Department of Physics , University of Maryland, College Park, MD, 20742\\
$^{2}$Department of Astronomy, University of Maryland, College Park, MD, 20742\\
$^{3}$Joint Space-Science Institute, College Park, MD, 20742
}

\begin{document}
\label{firstpage}
\pagerange{\pageref{firstpage}--\pageref{lastpage}}
\maketitle

\begin{abstract}
The possible {\it Fermi} detection of an electromagnetic counterpart to the double black hole merger GW150914 has inspired many theoretical models, some of which propose that the holes spiraled together inside a massive star. However, we show that the heat produced by the dynamical friction on such black hole orbits can exceed the stellar binding energy by a large factor, which means that this heat could destroy the star and thus make it difficult for enough gas to be near the holes at merger to produce detectable photons. These considerations must be taken into account when models are proposed for electromagnetic counterparts to the coalescence of two stellar-mass black holes. We find that only when the two black holes form very close to the center can the star avoid destruction. In that case, dynamical friction can make the black holes coalesce faster than they would in vacuum, which leads to a modification of the gravitational waveform that is potentially observable by advanced LIGO. 
\end{abstract}
\maketitle

\begin{keywords}
black hole physics -- gravitational waves -- stars: black holes -- (stars:) gamma-ray burst: general -- stars: massive
\end{keywords}

\section{Introduction}
 
The LIGO event GW150914 resulted from the coalescence of two black holes of masses of $35^{+5}_{-4} \ M_\odot$ and $29^{+4}_{-4} \ M_\odot$ \citep{Abbott16a}. The gravitational wave signal rose in frequency from 35~Hz to 150~Hz in $\sim 0.2$s. Just 0.4 seconds later, a signal was detected with the Fermi Gamma-ray Burst Monitor \citep{Connaughton16}. The signal lasted $\sim 1$~s and the isotropic equivalent luminosity of the non-thermal X-ray component was $\sim 10^{49}~{\rm erg~s}^{-1}$ if the source was at the distance of GW150914.  In some respects, the signal was similar to short gamma-ray bursts (GRBs). However, doubts have been raised about the association of the electromagnetic (EM) signal with GW150914 \citep{Greiner16}, especially given that INTEGRAL did not detect the GRB \citep{Savchenko16}. LIGO later reported one more clear gravitational wave (GW) detection and a possible detection \citep{Abbott16b}, but no candidate EM counterparts were found in those events \citep{Abbott16c, Racusin16, Smartt16}. 

If GW150914 indeed had an EM counterpart, then it means that contrary to previous expectations, the two black holes could not have merged in a near-vacuum, unless they are charged \citep{Zhang16, Liebling16, Fraschetti16}. Independent of whether the GW150914--Fermi GRB association is real, it is interesting to explore models that allow the production of a GRB or any EM counterpart during black hole mergers. For example, in order to explain a GRB near the time of the black hole merger, \citet{Loeb16} proposed a model in which two black holes form via a bar instability inside a collapsing, rapidly rotating massive star. However, \citet{Woosley16} showed that only inside a star with extreme low metallicity and no mass loss can two black holes of the desired masses form. The jet production in this model is similar to the ``collapsar'' model for long GRBs \citep{Woosley93, MacFadyen99}, except that an accretion--jet system forms around the binary black holes instead of forming from a single black hole. \citet{Woosley16} and \citet{Janiuk17} also proposed a second model: in a binary stellar system of two massive stars, the more massive star collapses to a black hole first and then enters the envelope of the other star. Eventually the core of the second star also collapses to a black hole, and an accretion disk--jet system forms around the binary black holes. There are also other models which involve a pre-existing accretion disk \citep{Bartos16, Murase16, Perna16, Stone16} to explain an EM counterpart in double black hole mergers.

Several concerns have been expressed regarding these models.  For example, \citet{Kimura16} perform a detailed calculation to explore the model of \citet{Perna16}, and find that the mass of the dead disk needs to be much greater than the mass of the disk proposed in \citet{Perna16} to explain the accretion rate and observed GRB luminosity. As another example, \citet{Lyutikov16} find that the magnetic flux needed to trigger such a jet is as high as $10^{12}$ Gauss and is difficult to form in such environments. 

In this paper, we raise a concern which has been discussed extensively in the context of common envelopes (see the review by \citealt{Ivanova13}) but so far has not received much attention in EM counterpart analyses. When two black holes orbit within a star, the heat produced by dynamical friction can eject most of the stellar material. If this happens, there may not be enough material to form a GRB once the two black holes merge, and thus the coalescence will be similar to a merger in vacuum. On the other hand, if dense stellar material is still around at merger, then the gravitational waveform can be different from that in vacuum. In the future if dual EM--GW signals are observed from double stellar-mass black hole mergers, then one should need to include these considerations when proposing a stellar model to explain the EM counterpart.

In Section 2, we use a simple stellar model to calculate the heat generated by dynamical friction when two black holes spiral inside a star. We compare this heat with the binding energy of the star, and find that in most scenarios the injected heat is many times the stellar binding energy. In Section 3, we calculate the coalescence time of the black holes, including the effect of dynamical friction. We find that the gravitational radiation waveform can be modified from that in vacuum. In Section 4, we summarize and discuss our results.

\section{Black holes orbiting each other inside a star}

As the two black holes orbit each other inside a star, dynamical friction and gravitational radiation reduce the separation between the black holes until they eventually merge. If dynamical friction dominates the inspiral, then most of the gravitational binding energy between the black holes is converted to heat. As we show in Section~2.1, the released gravitational energy can be many times greater than the self--binding energy of the star, which means that the inspiral has the potential to destroy the star. 

We focus on dynamical friction because we believe that the energy released in this way is understood more robustly than the energy release from accretion onto the holes, which we do not include in our analysis.  In a qualitative sense, the reason for the uncertainty in the accretion energy release involves two factors.  First, as has been discussed in many papers on accretion, the rate at which gas reaches a few gravitational radii from the holes (where the energy release is greatest) depends on highly uncertain physics involving winds and other processes (e.g., \citealt{Miller15}).  Second, at the very super-Eddington accretion rates that would be expected in a dense gaseous environment such as a stellar interior, photon trapping could lead to extremely radiatively inefficient flows \citep{Begelman78, Abramowicz88}, although the efficiency of such accretion is still under investigation \citep{Ohsuga05, Jiang14, McKinney15, Sadowski16}.

The injection of energy due to dynamical friction occurs on a length scale $R_{\rm DF}\sim GM/(v^2+c_s^2)$, where $M$ is the black hole mass, $G$ is the gravitational constant, $c_s$ is the sound speed of the gas, and $v$ is the net speed of the hole relative to the gas.  This is a far larger scale than the gravitational radius $R_g=GM/c^2$; indeed, if the stellar mass interior to the black hole orbit is less than or comparable to the black hole mass then $R_{\rm DF}$ is comparable to the orbital radius.  We do note that this same comparison of radii means that if most of the matter reaches the black holes and the resulting accretion is at least moderately radiatively efficient, then accretion could contribute significantly to the overall energy budget. Therefore, by only including the heat generated by dynamical friction, we get a conservative estimate of whether the star is disrupted. 

There are several models that can lead to two black holes orbiting inside a star starting from different initial separations. If the black holes are initially at a larger distance from each other, then they can orbit inside the star for a longer time, so more heat is injected to the star. It is intuitive that the star is likely to be destroyed when the black holes enter the star and start orbiting from near the surface. In contrast, the star has the best chance to survive when the black holes are formed very close to the center of the star. Therefore, we shall start from the close-origin scenario, and will carry out a quantitative analysis to show that the star can be destroyed even in this case.

\subsection{Two black holes formed within one massive star}
\begin{figure}
\centering
\includegraphics[width=3.1in]{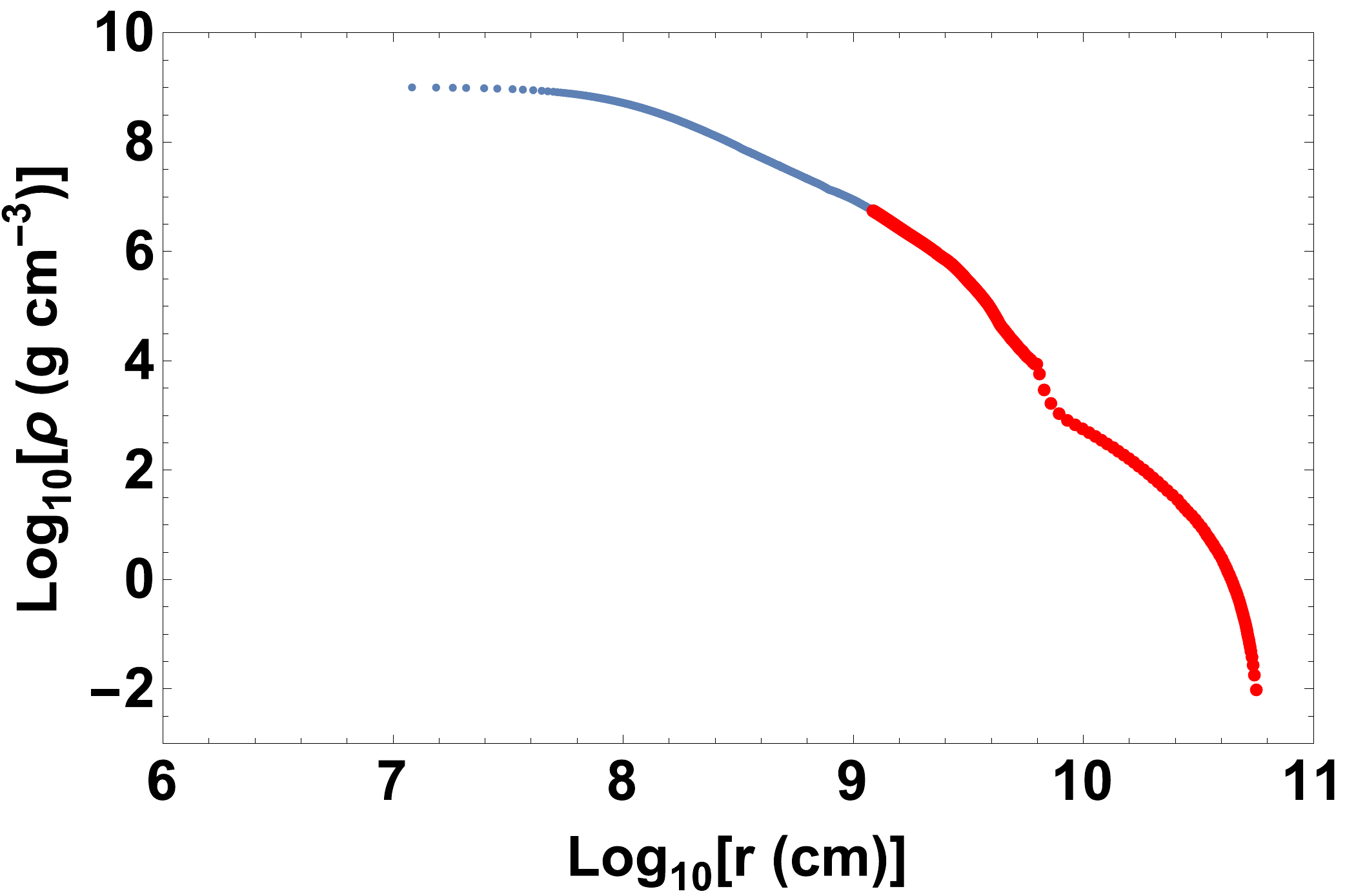}
\caption{The density--radius plot of R150A when the central density reaches $\sim 10^9$~g~cm$^{-3}$, before core collapse is about to start. The blue points show the inner portion of the star with a mass of $60~M_\odot$, which can later collapse to two $30 M_\odot$ black holes. The red points show the outer portion of the star with a mass of $90~M_\odot$. Data kindly provided by S. Woosley.}
\label{figureR150}
\end{figure}

\citet{Woosley16} showed that when a fast-rotating massive star goes through chemically homogeneous evolution with no mass loss, it is possible to form two black holes in the center due to bifurcation of angular momentum in the disk during core collapse \citep{Fryer01, Reisswig13}. An example of a successful evolution is provided by his Model R150A, in which the star has an initial mass of $150 M_\odot$ in the main-sequence phase. Core collapse starts when the central density reaches $\sim 10^9$~g~cm$^{-3}$. A snapshot of the density--radius profile at this stage is shown in Fig. \ref{figureR150}. As the two black holes are formed in the center, and the gravitational binding and thermal energy of the star are only $\sim1\%$ of the rest-mass energy of the star, the initial inner $60~M_\odot$ of the star (marked by blue dots) will correspond to the part which collapses to form two $30~M_\odot$ black holes. The outer $90~M_\odot$ of the star (marked by red dots) will redistribute as core collapse continues and black holes accrete and spiral towards each other.

Inspired by the R150 model, we consider a star with an initial mass of $150 M_\odot$.  As the central part of the star collapses, we suppose that two black holes, each of mass $30 M_\odot$, form at an initial distance $R_I$ from the center of the star. As the two black holes spiral in, the core continues to collapse and accrete onto the black holes, so the density profile in the collapsing region will become different from Fig. \ref{figureR150}. \citet{Woosley95} show that when a massive star is in the pre-supernova phase, the density in the collapsing central region is approximately a power law function of radius $\rho\propto r^{-\gamma}$ with an index up to $\gamma\sim 2.5$. When an accretion disk forms around the central black hole, the disk region has a density that is also a power-law function of radius, but with an index that is more likely to be $\gamma\sim 1.5$ \citep{Popham99, MacFadyen99}. Here we simplify by assuming that the stellar density profile in the central region where the black holes coalesce is a single, time-independent, power law. If the central density at $R_c$ is $\rho_c$, then
\bey
\rho (r) = \rho_c \left( \frac{r}{R_c} \right)^{-\gamma},
\eey
where $R_c$ is taken to be the gravitational radius of a $30~M_\odot$ black hole, which is $4.4\times10^6$~cm. 

We wish to determine whether the star can withstand the heat produced during the binary black hole coalescence, as a function of $\rho_c$, $\gamma$, and $R_I$. The stellar material needs to be dense enough to form black holes, so we consider a central density range $10^8 \rm{\ g \ cm}^{-3} \leq \rho_{\rm c} \leq 10^{12}$g cm$^{-3}$ \citep{Popham99}. $R_I$ must be larger than $R_c = 4.4 \times 10^6$ cm.  We could also impose an upper bound on $R_I$ because the black holes cannot be formed in a region that has too low a density, but as we show below the larger $R_I$ is, the more energy is injected into the star via dynamical friction.  Thus the star has the greatest likelihood of survival if $R_I$ is as small as possible.  We allow $\gamma$ to vary between 1.5 (Bondi free-falling profile) and 2.5 (pre-supernova phase profile).

\begin{figure}
\centering
\includegraphics[width=3.1in]{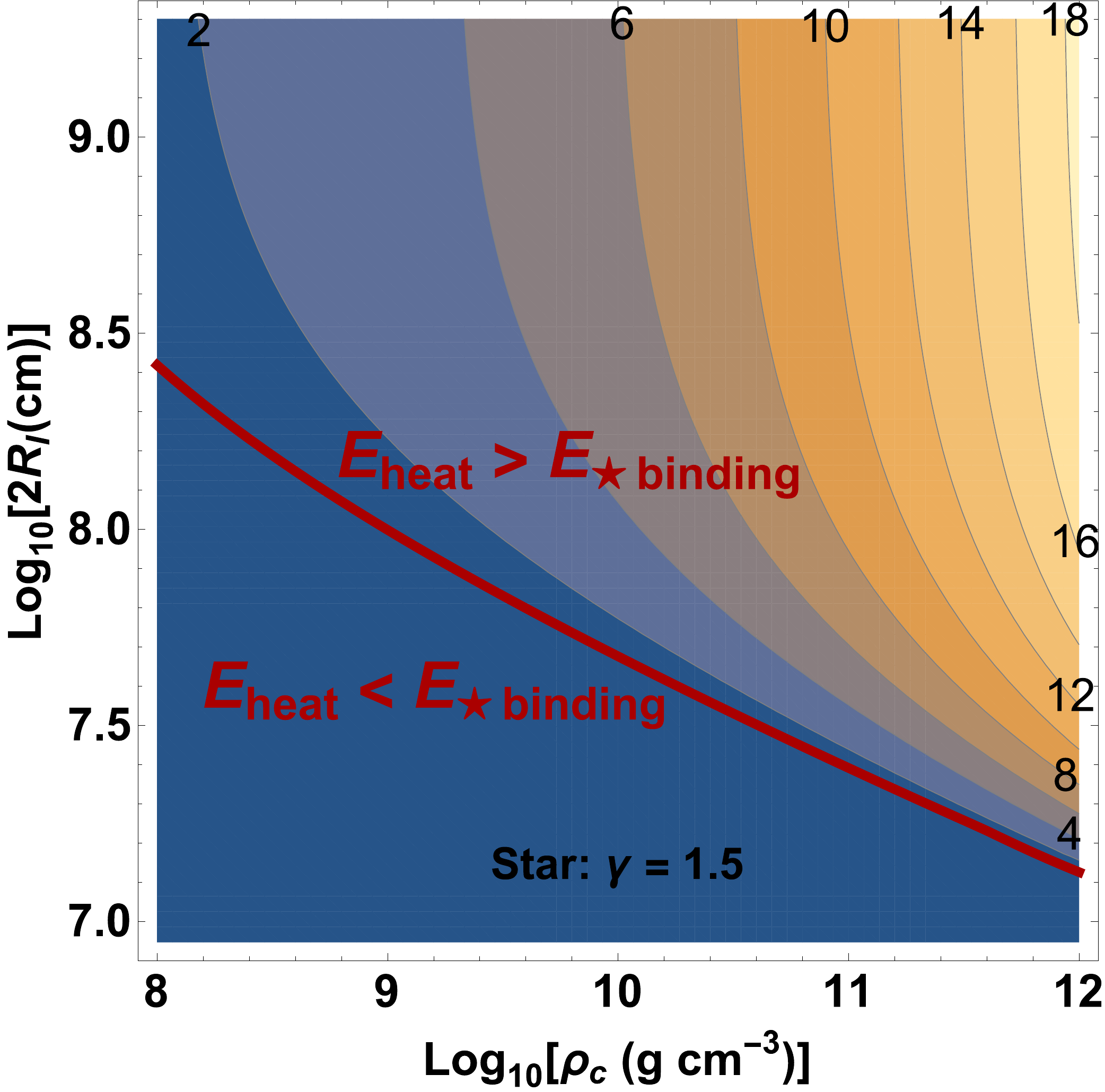}
\caption{Contour plot of the ratio of total amount of dynamical-friction-generated heat to the stellar binding energy, during the coalescence of two $30 M_\odot$ black holes within the remaining $90 M_\odot$ stellar material after core collapse of a $150 M_\odot$ star. The x-axis is the stellar central density, and the y-axis is the initial separation between the two black holes. The stellar density in the central region goes as $\rho(r) \propto r^{-1.5}$. The value labeled on the contour is $E_{\rm heat} / E_{\rm B}$. The red curve shows where $E_{\rm heat} = E_{\rm B}$. Above it, the heat is many times greater than the stellar binding energy, so it is hard for the star to withstand the heat and stay intact. This figure shows that only when the two black holes are formed very close to each other can the star survive the heat produced until merger. Otherwise, most of the stellar material is ejected and there may not be enough gas left to produce a detectable EM signal at merger.}
\label{model1}
\end{figure}

For the initial black hole separations of greatest interest, the gas mass within the black hole orbit is small compared to the masses of the black holes. In this situation \citet{Escala04} show that the gas close to the binary forms an ellipsoid.  The resulting tidal force between the ellipsoid and the outer spherical gas removes the orbital energy of the binary with an efficiency comparable to that of dynamical friction. Therefore we assume that for each black hole the heat is produced at the rate $P_{\rm DF}$ given by dynamical friction when the orbital speed $v_{\rm BH}$ is much greater than the sound speed $c_s$:
\bey
P_{\rm DF} = 4\pi\rho (GM_{\rm BH})^2v_{\rm BH}^{-1}
\eey
\citep{Chandrasekhar43, Ostriker99}, where $\rho$ is the density of the ambient medium, and $M_{\rm BH}$ is the mass of the black hole.

The maximum radiation power that can escape from the surface of the star is the Eddington luminosity $L_{\rm Edd}\approx 2\times 10^{40}~(M/150~M_\odot)$~erg~$s^{-1}$. When each black hole is at a distance $r$ from the center, the instantaneous dynamical friction power is:
\bey
P_{\rm DF} = 10^{49.4+2.4\times(2.5-\gamma)}  \frac{\rho_ c} {10^8  \rm {g \ cm}^{-3}}  \left( \frac{r}{10^9 {\rm cm}} \right)^{0.5-\gamma} {\rm erg \ s}^{-1}.
\eey
$P_{\rm DF} $ is at least nine orders of magnitude above the Eddington luminosity, so heat cannot escape from the stellar envelope through radiative diffusion. Instead, the heat is expected to eject stellar mass.  It is therefore important to determine whether the total injected energy exceeds the self binding energy of the star.  If it does, then the star could be destroyed. We note that in a standard common envelope scenario, the efficiency of converting black hole gravitational binding energy to eject gas is usually assumed to be $\lesssim 0.5$. However, our system is embedded deeply inside dense stellar material instead of a tenuous envelope, so the time needed for the energy to escape is much longer than the time needed for the holes to coalesce.

\begin{figure}
\centering
\includegraphics[width=3.1in]{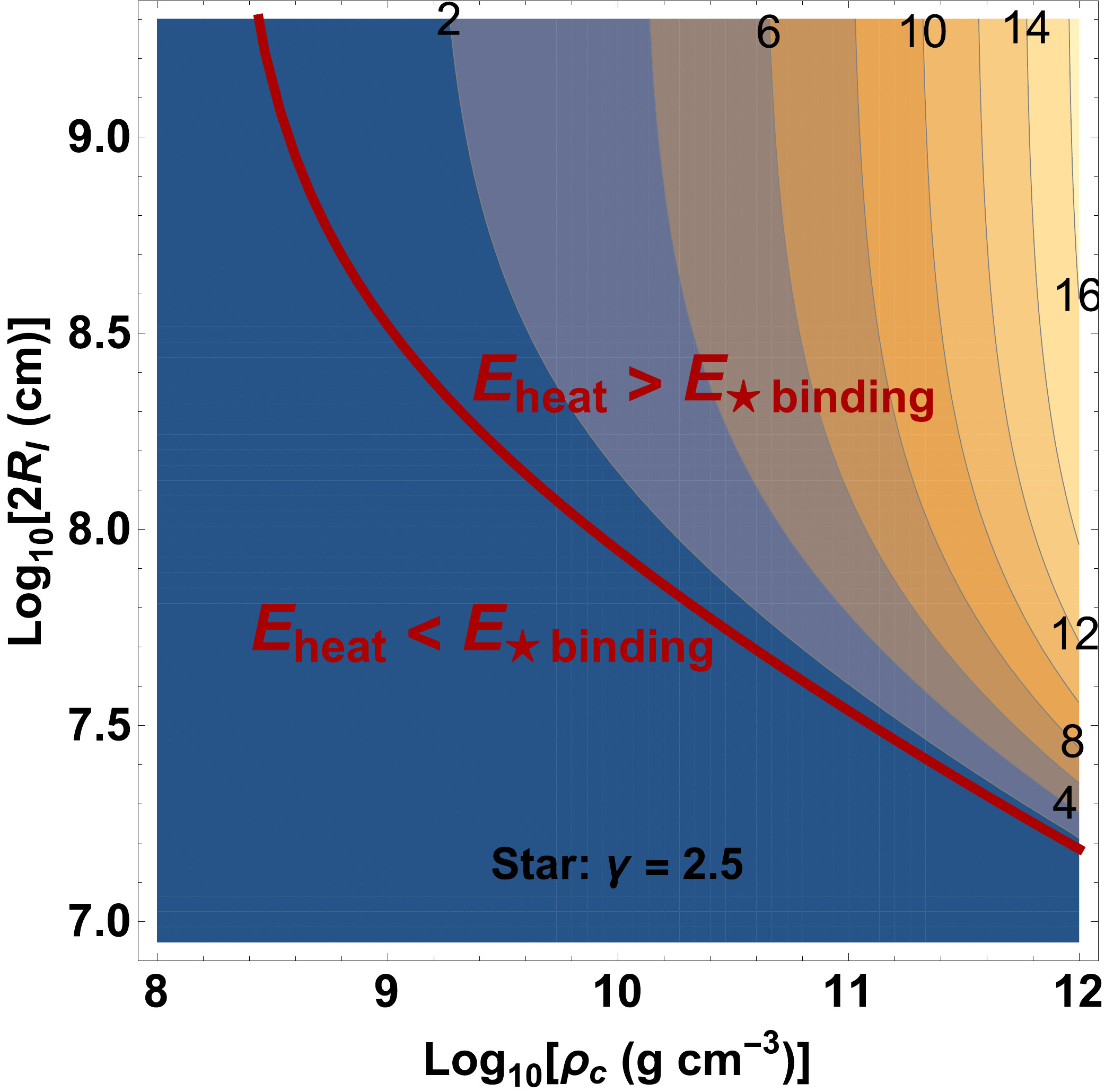}
\caption{The same plot as Fig. \ref{model1}, except that we use a model of $\rho(r) \propto r^{-2.5}$ for the stellar central region.}
\label{model2}
\end{figure}

The total injected heat can be calculated by integrating the dynamical friction power along the path of the black hole inspiral. The black hole inspiral path depends on whether gravitational radiation dominates or dynamical friction dominates. As the gravitational radiation power depends strongly on the separation, $P_{\rm GR}\propto r^{-5}$, whereas the dynamical friction power $P_{\rm DF} \propto r^{0.5-\gamma}$ has a much weaker dependence on radius, there is a distance $R_{\rm eq}$ at which the two contribute equally to the inspiral. When the black hole separation is $r>R_{\rm eq}$, dynamical friction dominates and we assume that the injected heat is simply the change in the black hole gravitational binding energy as the orbit shrinks.  In contrast, when the orbital radius $r<R_{\rm eq}$, gravitational radiation takes over and the path of the black holes is almost a gravitational inspiral in vacuum. Thus in this phase we compute the injected heat by integrating the dynamical friction power along the gravitational inspiral path from $R_{\rm eq}$ until merger. Denser gas around the black holes and a larger initial separation naturally lead to greater heat production.

We calculate the gravitational binding energy of the outer $90~M_\odot$ of the star before core collapse happens, using a curve fitted to the density--radius profile (the red part) in Fig. \ref{figureR150} . Heat produced during core collapse and black hole accretion can only be injected into the remaining stellar material and make it less bound. Moreover, neutrino production is negligible in the cool outer stellar material, so energy cannot escape efficiently. Therefore, the binding energy calculated this way is an upper bound to the stellar binding energy at later phases:
\bey
E_B \leq \int \frac{G (m(r) +60 M_\odot)}{r} 4 \pi r^2 \rho(r) dr \approx 9.9\times10^{53}~{\rm erg}.
\eey
If $E_{\rm heat}$, the total amount of heat produced by dynamical friction from $R_I$ until merger, exceeds this maximum stellar binding energy, then the star will almost certainly be disrupted. 

We show the ratio of $E_{\rm heat}$ to  $E_{\rm B}$ in Fig. \ref{model1} for $\gamma=1.5$. It is clear that only when the two black holes form very close to each other (e.g., $R_I \lesssim10^{8}$ cm when $\rho_c \geq 10^9 \rm{g~cm^{-3}}$) can the star survive the heat that is injected. Fig. \ref{model2} shows that for a steeper density profile $\gamma = 2.5$, the condition for the star to remain bound is slightly relaxed (e.g., $R_I \lesssim10^{8.5}$ cm when $\rho_c \geq 10^9 \rm{g~cm^{-3}}$). If the two black holes form too far from each other, then the heat accumulated during inspiral is likely to unbind the star, and there may not be enough material left at merger to produce detectable EM signals.

\subsection{Other scenarios}

We have shown that the energy injected by dynamical friction is sufficient to destroy the star unless the black holes are formed extremely close to each other.  It therefore follows that if the black holes are created {\it outside} the star, the star is extremely susceptible to destruction.  Thus energy injection must be considered in all such scenarios.  As an example of a model in this class, \citet{Woosley16} proposed that in a massive binary system, one star can evolve into a black hole and then spiral into the other star and, during the inspiral, the core of the second star can collapse to a black hole.  Another possibility would be a triple system in which two massive stars collapse to black holes and are then enveloped by the third star when it becomes a giant.

It is likely that the nature of gas ejection will differ between specific scenarios.  If the black holes are formed deep within the star then the injected energy might not have time to escape before the holes merge.  This could result in an explosion that ejects the whole envelope.  If instead the holes are created outside the star, it seems more likely that the stellar envelope will be gradually peeled off and thus that matter will be continuously unbound.  

\vspace{-0.5cm}

\section{Inspiral time inside a star}

When two black holes spiral inside a star instead of in vacuum, their coalescence time is shorter than the gravitational inspiral timescale, because dynamical friction also shrinks their orbit. As the GW frequency is twice the orbital frequency of the binary black holes, the GW signal will sweep from low frequency to high frequency faster than in vacuum. 

\begin{figure}
\centering
\includegraphics[width=3.1in]{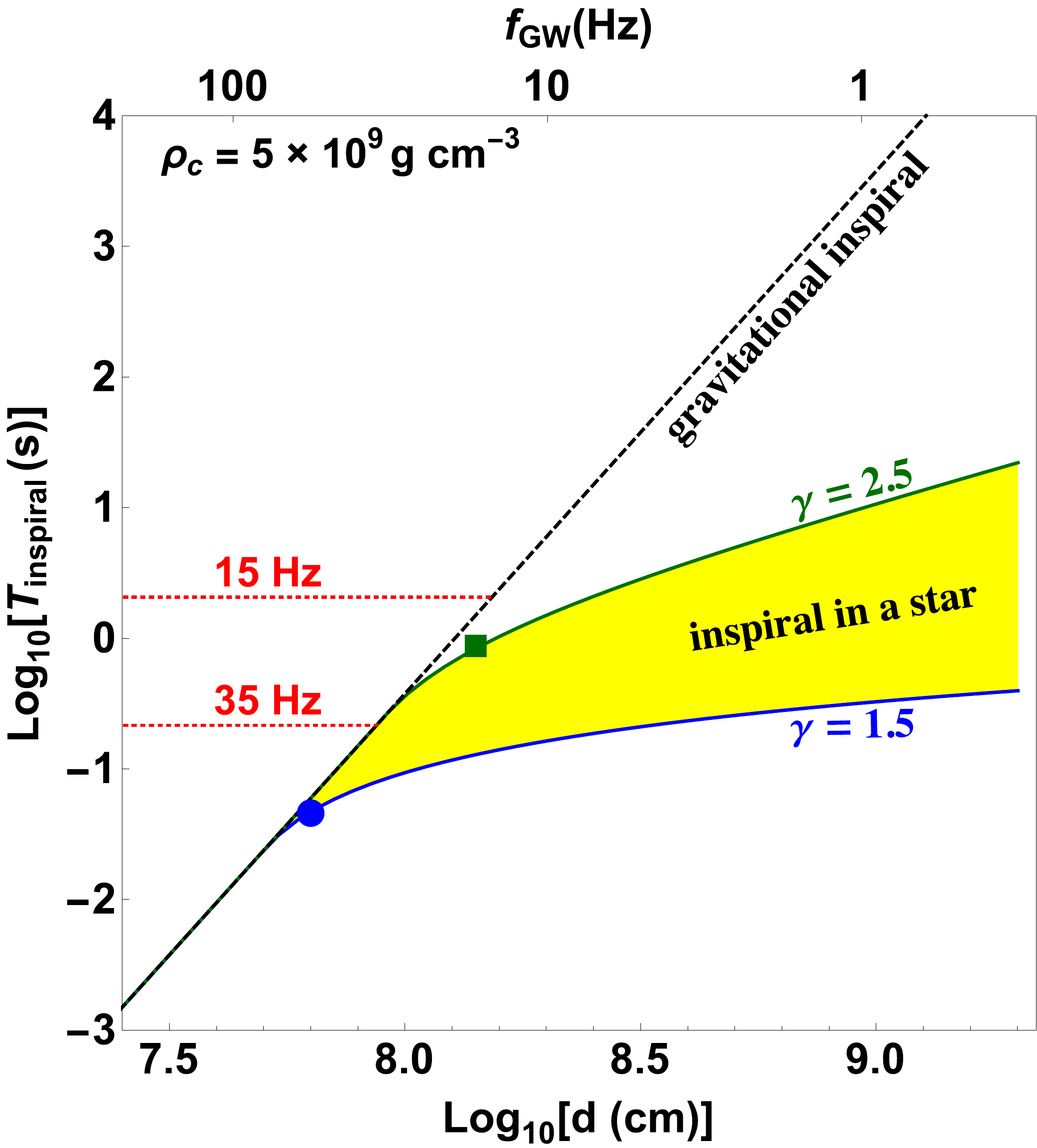}
\caption{The remaining coalescence time (y-axis) of two 30 $M_\odot$ black holes inside the central stellar region with $\rho_c = 5\times10^{9} \rm \ {g \ cm^{-3}}$, as a function of the distance $d$ between two black holes (lower x-axis), or the corresponding gravitational wave frequency (upper x-axis). The black dashed line shows a pure gravitational inspiral in vacuum for comparison. The yellow-shaded region shows the inspiral within the star, taking account of how dynamical friction speeds up coalescence but assuming the star remains intact against heat generated by dynamical friction. The blue curve uses a model with a density index $\gamma =1.5$ for the central region, and the blue filled circle indicates the initial distance between black holes which gives $E_{\rm heat} = E_B$. The green curve uses a model with $\gamma = 2.5$ for the central region, and the green filled square is the initial distance which gives $E_{\rm heat} = E_B$. We note that if the black holes are formed to the right of blue circle and the green square, the star is likely to be disrupted by the heat accumulated before merger. The two red dotted lines indicate the current and future LIGO sensitive low-frequency bands.}
\label{inspiral}
\end{figure}

If the black holes form close enough so that the star stays intact until merger, we can observe modifications of the GW signal, as we show in Fig. \ref{inspiral}.  Therefore, GW signals can potentially disclose whether a merger happens in vacuum or in a star, and can be used to constrain the stellar parameters. For a star with $\rho_c\sim10^9~\rm{g~cm^{-3}}$, it will be hard to tell whether the $30 M_\odot$ black holes merge inside a star or in vacuum with the current LIGO sensitivity. However, when ground-based detectors become more sensitive at lower frequencies, we will be able to see deviations between an inspiral inside a star and a gravitational inspiral. For a star with $\rho_c\sim10^{10}~\rm{g~cm^{-3}}$ or higher, advanced LIGO (aLIGO) can certainly tell if the inspiral is in vacuum or in a star. 

\vspace{-0.5cm}

\section{Discussion}

When two black holes orbit within a star, they have the potential to produce joint EM--GW signals. Here we point out that the heat produced because of dynamical friction between the black holes and the stellar medium is energetically sufficient to eject all of the gas from the system, which would therefore prevent the production of detectable EM signals close to merger.  Only if the holes are formed so close to each other that their inspiral is dominated by gravitational radiation might the heat input be small enough that the star could survive.  These considerations apply to any model that uses a star to provide material for the EM signal, e.g., the formation of two black holes inside one massive star, or one black hole entering another star (and eventually merging with the black hole collapsed from the core of the star), or two black holes entering a third star. 

For our model we assume the gas near the center stays unaffected by dynamical friction. It is true that the injection of energy via dynamical friction will itself change the properties of the nearby gas. Indeed, if there were no surrounding stellar envelope, the expected effect would be to heat up the gas and reduce its density, which might lead to a self-regulation of the production of heat by dynamical friction. However, in the situation that we envision and that has been proposed in several models, the rest of the star provides a large overburden of mass on top of the black hole binary.  As a result, if the density of the gas near the orbit is decreased significantly due to the injection of energy, we expect that the system will become Rayleigh-Taylor unstable and thus mass will flow into the orbital region on a dynamical timescale.  It is difficult to determine the time-averaged state of such a system without very detailed computations, but we anticipate that the net result will be that the rate of energy injected by dynamical friction will be at most a factor of a few less than the rate computed using the unperturbed stellar structure. Given that for most initial separations we find that the injected energy can easily be many times greater than the stellar binding energy, we do not expect that a slight reduction in the time-averaged gas density near the holes will change our conclusions.

We emphasize that the destruction of the star is not necessarily absolute. If the coalescence time is much shorter than the sound crossing time of the stellar core then the star may avoid disruption before merger. Whether or not a GRB can form in such a short amount of time is worth investigating but beyond the scope of this paper. Furthermore, we have a complex system, so there can be ways that a small amount of matter can remain in the system. For example, if the gas is Rayleigh-Taylor unstable, then it is conceivable that some mass could reach the holes in dense filaments, although some simulations in the possibly comparable context of Bondi-Hoyle accretion onto black holes find that such filaments are easily evaporated \citep[e.g.,][]{Park11, Park12}. Focused high-resolution simulations in the stellar context would be needed to address this question. For the putative Fermi counterpart to GW150914, only $\sim 10^{-4}~M_\odot$ of gas is required for the observed EM power \citep{Connaughton16}. It will be difficult to rule out such a small retention fraction. However, any model that proposes an EM counterpart due to inspiral inside a star will need to make a clear case that significant matter can remain bound despite the large amount of energy that is injected due to dynamical friction.

If the two black holes in a merger event are formed inside a star, there could be unique observational features. For example, if the black holes are formed relatively far away from each other and thus the heat generated during their inspiral is orders of magnitude higher than the stellar binding energy, the result could be a violent explosion. The energy scale of the explosion would be comparable to a supernova. However, given that negligible $^{56}{\rm Ni}$ might be produced in this process, the light curve of this optical transient would look very different from a supernova. On the other hand, if the two black holes are formed very close to each other and the star can stay intact until merger, dynamical friction will make the black holes coalesce faster than they would in vacuum. Therefore, the GW signal emitted before merger can have a different waveform than that of a gravitational inspiral in vacuum. This effect could be detected with ground-based instruments (such as aLIGO) with improved sensitivity at low frequencies, and should be included in the templates used to search for GW signals from double black hole mergers. When such modified gravitational waveforms are detected in the future, one can more efficiently conduct a search for the associated EM signals.

In summary, we find that the heat injection due to dynamical friction poses a major challenge to any models of EM counterparts to double stellar-mass black hole mergers that involve a surrounding stellar envelope. If such scenarios do play out in nature, they can lead to new observational signatures in both the electromagnetic and gravitational wave domains.

\vspace{-0.5cm}
\section*{Acknowledgements}
We thank Stan Woosley for fruitful discussions and kindly providing data of R150A model. We also thank Ilya Mandel and Brian Metzger for helpful comments. L.D. and J.C.M. acknowledge NASA/NSF/TCAN (NNX14AB46G), NSF/XSEDE/TACC (TG-PHY120005 and TG-AST160003), and NASA/Pleiades (SMD-14-5451). M.C.M. acknowledges NSF (AST-1333514).

\vspace{-0.5cm}

\input{DF_v11.bbl}
\bsp	
\label{lastpage}
\end{document}